



\def\={\!=\!}
\def\-{\!-\!}
\def\a{\alpha}

\def\d{\partial}

\def\da{^{\dagger}}
\def\e{\eqno}

\def\ie{{\it i.e.}}
\def\pg{paragrassmann}
\def\q{\quad}
\def\qq{\qquad}
\def\qm{quantum mechanics}
\def\ss{supersymmetric}
\def\t{\theta}

\def\z{Z}
\def\1{{\textstyle{1\over2}}}

\magnification=1200
\vsize=22truecm
\hsize=15truecm
\voffset=0.5truecm
\hoffset=1truecm
\baselineskip=18truept
\lineskip=1pt
\lineskiplimit=0pt
\parskip=6truept


\font\titre=cmbx10 scaled\magstep2
\font\bigletter=cmr10 scaled\magstep2


\baselineskip=12pt

\line{\hfill McGill/92-54}
\line{\hfill hep-th/9305128}
\line{\hfill April 1993}

\vskip 1in
\centerline {\titre Fractional Supersymmetry}
\vskip 0.03in
\centerline {\titre and}
\vskip 0.03in
\centerline {\titre Quantum Mechanics}
\vskip 0.75in
\centerline{{{\bigletter S}T\'EPHANE {\bigletter D}URAND}\footnote{$^{*}$}
{E-mail address: durand@hep.physics.mcgill.ca}}
\vskip 0.2in
\centerline{\it Department of Physics}
\centerline{\it McGill University}
\centerline{\it 3600 University Street}
\centerline{\it Montr\'eal, PQ, H3A 2T8, Canada}
\vskip 0.45in

\centerline{To appear in {\it Phys. Lett.} {\bf B312}, 115 (1993).}
\vskip 0.45in

\centerline{\bf Abstract}
\vskip 0.1in

\noindent
We present a set of quantum-mechanical Hamiltonians
which can be written as the
$F^{\,\rm th}$ power of a conserved charge: $H=Q^F$ with $[H,Q]=0$ and
$F=2,3,...\, .$
This new construction, which we call {\it fractional}\/ supersymmetric quantum
mechanics, is realized in terms of \pg\ variables satisfying $\t^F=0$.
Furthermore, in a pseudo-classical context, we describe {\it fractional}\/
supersymmetry transformations as the $F^{\,\rm th}$ roots of time translations,
and provide an action invariant under such transformations.

\vfill
\eject

\baselineskip=18truept

\noindent {\bf 1. Introduction}
\vskip 0.2cm

Supersymmetric (SUSY) quantum mechanics has applications in nuclear and atomic
physics, and is also a simpler framework to understand new ideas before
extending them into SUSY field theories.
In SUSY quantum mechanics, the Hamiltonian $H$ is written as the square of a
conserved supercharge: $Q^2=H$. In its simplest realization, it describes a
spin-1/2 particle.
Recently, the para-supersymmetric (PSUSY) generalization of SUSY quantum
mechanics was developed$^{[1,2]}$. A PSUSY quantum-mechanical Hamiltonian of
order 3, which describes a spin-1 particle, is realized as$^{[1]}$: $Q^3=QH$
with $Q^2\!\not=\!H$ (where $Q$ is now a conserved para-supercharge). In
general, a PSUSY quantum-mechanical Hamiltonian of order $M$ is of the form
$Q^M=Q^{M-2}H+...$ and describes a spin-$(M-1)/2$ particle$^{[2]}$.

In this letter, we present an alternative generalization of SUSY quantum
mechanics which we call
{\it fractional}\/ supersymmetric (FSUSY) quantum mechanics of order $F$.
Hamiltonians of this new kind are expressed as the $F^{\,\rm th}$ power of a
conserved {\it fractional}\/ supercharge: $Q^F=H$ with $F=2,3,...\,$.
This means that the FSUSY transformations are the $F^{\,\rm th}$ roots of time
translation. We discuss these transformations in a {\it pseudo-classical}\/
Lagrangian context, and provide a FSUSY invariant action (a
fractional-superspace formulation is given in Ref.~[3]).

The internal space of the quantum-mechanical systems is described by a \pg\
variable $\t$ of order $F$ satisfying $\t^F=0$. In matrix form, the
Hamiltonians are
realized in terms of $F\!\times\!F$ matrices. The spectrum of the
one-dimensional harmonic oscillator (or equivalently the two-dimensional
constant magnetic field) is found to be $F$-fold degenerate above the ground
state, which can be both {\it unique}\/ and {\it degenerate}\/. We also
describe a conformal symmetry, namely dilations, as a dynamical
symmetry for the $1/x^2$ potential. In Ref.~[4], an alternative realization of
the FSUSY algebra is given in which $q$-deformed relations are found among
different conserved charges.

\vskip 0.5cm
\noindent {\bf 2. Paragrassmann variables}
\vskip 0.2cm

In this section, we introduce generalized variables which
interpolate between ordinary bosonic and fermionic ones.
They can be interpreted either as generalized coordinates,
or generalized creation and annihilation operators. The
latter interpretation is more relevant in the present
quantum-mechanical context, but the former
point of view is used in Ref.~[3]
when discussing the fractional superspace formulation of
FSUSY transformations.
The notation, however, will be more reminiscent of the generalized
coordinates interpretation. A more complete presentation of this formalism is
given in Ref.~[4].

We introduce a paragrassmann variable $\t$ of order $F$, and
its derivative $\d\equiv \d/\d\t$, which satisfy
$$\eqalign{\t^F=0&, \qq \d^F=0, \qq F=1,2,...   \cr
          &(\t^{F-1}\not=0, \; \d^{F-1}\not=0). \cr} \eqno(1)$$
In order to be able to recover the 3 different limits which we describe below
(fermionic, bosonic and ``null"), we take the
generalized commutation relation between $\t$ and $\d$ to be
$$[\d,\t]_q\equiv\d\t-q\t\d=\a(1-q),      \eqno(2)$$
where $\a$ is a free parameter and $q\in\cal C$ a {\it primitive}\/
$F^{\,\rm th}$ root of unity:
$$q^F=1 \qq (q^n\not=1 ~~{\rm for}~~ 0<n<F).  \eqno(3)$$
By a {\it primitive}\/ root, we mean a root satifying the
condition in parentheses; for instance, $q\not=\pm 1$ for $F=4$.
[We will see below that the
condition $(3)$ is actually a consequence of $(1)$ and $(2)$.]
First, note that the null limit $F=1$ $(q=1)$, that is $\t=\d=0$, is
well-defined since the r.h.s of (2) is zero for $q=1$ (and for finite $\a$).
In previous works$^{[5,6]}$ on \pg\ variables, the r.h.s of $(2)$ was chosen to
be 1 instead of $\a(1-q)$, which is inconsistent in the limit $F=1$
(which we use in Ref.~[3,4]).
Second, note that we recover
the ordinary grassmann case $(q=-1)$ for $F=2$, \ie, $\{\d,\t\}=2\a$.
Third, for some choices of $\a$,
we also recover (within factors) the bosonic case $(q=1)$ for $F\to\infty$.
For instance,
for $\a=F$ we find that the r.h.s of (2) is finite and non-zero:
$$\lim_{F\to\infty}F(1-q)=-2\pi i.                   \eqno(4)$$
Strictly speaking, this is the bosonic limit for the first root
$q=exp(2\pi i/F)$.
In the context of a fractional superspace formalism$^{[3]}$, we can choose to
work with real $\t$ and $\d$, whereupon the consistency of the relation (2)
under hermitian conjugation implies that $\a$ must be real.
In the following sections, we will not be concerned with the $F\to\infty$
limit, so we let $\alpha$ remain unfixed.

The definition (2) implies
$$\d\cdot\t^n=\a(1-q^n)\,\t^{n-1}+q^n\t^n\d. \eqno(5)$$
Setting $n=F$ in $(5)$ demonstrates that the consistency of the formalism
requires the condition $(3)$.
We shall also need
the operator $B_{(F)}$ defined as
$$B_{(F)}=\sum_{i=0}^{\infty}c_i\t^i\d^i
=c_0+\sum_{i=1}^{F-1}c_i\t^i\d^i \eqno(6a)$$
with
$$c_0=(1-F)/2 \q {\rm and} \q c_i=[\a^i(1-q^i)]^{-1}
\q (i=1,2,...\, ,F-1). \eqno(6b)$$
This operator has the following properties:
$$[B_{(F)},\t]=\t, \qq [B_{(F)},\d]=-\d. \eqno(7)$$
One may also introduce other derivatives which satisfy $q$-deformed relations
between themselves (see Ref.~[4]).

A matrix realization of
$\t$ and $\d$ is given by
$$\t={\pmatrix{0&a_1&0&0&0\cr
               0&0&a_2&0&0\cr
               0&0&0&\ddots&0\cr
               0&0&0&0&a_{F-1}\cr
               0&0&0&0&0\cr}}, \qq
  \d={\pmatrix{0&0&0&0&0\cr
               b_1&0&0&0&0\cr
               0&b_2&0&0&0\cr
               0&0&\ddots&0&0\cr
               0&0&0&b_{F-1}&0\cr}} \e(8a)$$
with the constraint (no summation on $i$)
$$a_i b_i =\a(1-q^{-i}).     \eqno(8b)$$
Note that in general $\d\not=\t\da$.
In this matrix realization, $B_{(F)}$ is found to be the third component of the
spin-$(F-1)/2$ representation of the rotational group:
$$B_{(F)}=J_3^{[(F-1)/2]}. \eqno(9)$$

\vskip 0.5cm
\noindent {\bf 3. Fractional supersymmetric \qm}
\vskip 0.2cm

FSUSY \qm\ of order $F$ is defined through the algebra
$$Q^F=H, \qq [H,Q]=0, \qq F=2,3,... \e(10)$$
where $H$ is the Hamiltonian and $Q$ the conserved fractional supercharge.
The construction of $Q$ and $H$ will be realized in terms of \pg\ variables
$(\t,\d)$ of order $F$. Let us first work in one dimension.
We introduce the bosonic operators $a$ and $a\da$
$$a=[p+iW(x)]/\sqrt2, \qq a\da=[p-iW(x)] /\sqrt2         \eqno(11)$$
which satisfy
$$[a\da,a]={d\over dx}W(x)\equiv W'(x) \eqno(12)$$
where $p=-id/dx$.
A Hamiltonian and a conserved charge satisfying $(10)$ are given by:
$$Q=\d^{F-1}a+e P\t a\da+(1-P)\t \e(13)$$
$$H={1\over2}\left(p^2+W^2\right)+W'\cdot S \e(14)$$
with
$$S=-\textstyle{1\over2}+P, \qq P=e\t^{F-1}\d^{F-1}, \qq P^2=P \e(15)$$
and where the order-dependent constant $e$ is given by
$$e^{-1}=F\alpha^{F-1}.   \e(16)$$
In $(13)$ and $(15)$, it is understood that $\t$ and $\d$ satisfy the relations
(1-3) for a given order $F$.

For $F=2$, we have $P\t=\t$ and we recover for $Q$ and for $H$ (\ie, for $S$)
the ordinary \ss\ quantities:
$$Q=\d\,a+e\,\t a\da \eqno(17)$$
and
$$S=-\textstyle{1\over2}+e\,\t\d. \eqno(18)$$
With the realization $(8)$, we find
$$S=\textstyle{1\over2}\sigma_3
={1\over2}\pmatrix{1&0\cr0&-1\cr}.\eqno(19)$$
Thus the Hamiltonian $(14)$ may be interpreted as describing a ``spin-1/2"
particle
moving in a potential $W^2/2$ and a ``magnetic field" $W'$.
Moreover, with the choice $\a=1/2$ ($e=1$), we may choose $\t=\sigma_{+}$ and
$\d=\sigma_{-}$, \ie, $\d=\t\da$.

In the general case, using the realization $(8)$ in terms of $F\!\times\!F$
matrices, the projector $P$ is
found to be a matrix whose upper left entry is one and all others zero.
Therefore, the matrix $S$ is
$$S={1\over2}{\pmatrix{1&0&0&0&0\cr
              0&-1&0&0&0\cr
              0&0&-1&0&0\cr
              0&0&0&\ddots&0\cr
              0&0&0&0&-1\cr}}. \e(20)$$
Restricting ourself to the harmonic oscillator
potential, \ie\ $W=\pm\,\omega x$, the spectrum of the Hamiltonian $(14)$
is clearly
$$E=(n+{\textstyle{1\over2}}\pm s)\omega, \qq n=0,1,2,...  \e(21)$$
where $s$ stands for the eigenvalues of $S$, which are:
$s=+{1\over2}$ (non-degenerate) and $s=-{1\over2}$ ($F-1$ degenerate).
Unlike the SUSY case (and the
PSUSY$^{[1,2]}$ one; see below), the spectrum of the ``spin" matrix is not
invariant under $S\to-S$. Therefore the substitution $\omega\to-\omega$ leads
to a different spectrum; hence the $\pm$ sign in (21).
Let us explicitly compare the spectra for the SUSY, PSUSY of order 3, and FSUSY
of order 3 cases.
The energy levels are given by the formula $(21)$ with the following values for
$s$:
$$\eqalign{{\rm SUSY}: \q &s=\textstyle{1\over2},-\textstyle{1\over2}\cr
          {\rm PSUSY}: \q &s=1,0,-1\cr
          {\rm FSUSY}: \q &s=\textstyle{1\over2},
-\textstyle{1\over2},-\textstyle{1\over2}\cr}        \e(22)$$
The spectra are shown in Fig.~1.
The subscripts $(+)$ and $(-)$ refer to the two inequivalent possibilities
$\pm\,s$ of the FSUSY construction [{\it c.f.}\/ (21)].
In the SUSY and FSUSY$_{(-)}$ contexts, there is a unique ground state
with zero energy, with all other levels respectively 2-fold and 3-fold
degenerate. The only difference for the FSUSY$_{(+)}$ case is that the ground
state is 2-fold degenerate. In the PSUSY case, the unique ground state has
negative energy, the next level is 2-fold degenerate (and positive), and all
others
are 3-fold degenerate.

In general, we have the following for the FSUSY cases of arbitrary order.
With the ``minus" sign in (21), the spectrum is $F$-fold degenerate for all
levels
except for the {\it unique}\/ ground state (which has zero energy). Thus,
FSUSY is not spontaneously broken, \ie, the ground state is invariant under
FSUSY transformations since $Q|0\rangle=0$.
If this were not the case, one would have another state $|0\rangle'=Q|0\rangle$
with zero energy since $Q$ commutes with $H$.
On the other hand, with the ``plus" sign, the ground state still has zero
energy but is now $(F-1)$-fold degenerate,
and FSUSY might be spontaneously broken.
Like the SUSY case (but unlike the PSUSY case), the spectrum is non-negative.
Unlike the SUSY case, however, it is possible to have a degenerate vacuum with
zero energy. This is forbidden in SUSY since $H|0\rangle=0$ implies
$Q|0\rangle=0$. Indeed,
$0=\langle0|H|0\rangle=\langle0|Q^2|0\rangle=|Q|0\rangle|^2$ since $Q$ is
hermitian.

It is straightforward to extend our construction to purely magnetic
two-dimensional systems, \ie, to particles moving in the plane $(x,y)$ under
the influence of a magnetic field $B_z$ directed along the $z$-axis. Working
with the (gauge-independent) components $v_x=p_x-A_x$ and $v_y=p_y-A_y$ of the
velocity (where $A_x$ and $A_y$ are the two components of the potential
vector), we have $[v_x,v_y]=iB_z$. Therefore, to transpose the formulas (11-14)
from one- to two-dimensional systems, it suffices to make the substitutions
$p\to v_x$ and $W\to v_y$ which imply $W'\to-B_z$. The Hamiltonian is then
$$H={1\over2}(v_x^2+v_y^2)-B_z\!\cdot\!S \e(23)$$
which is a gauge-independent expression.

Let us now turn to the $1/x^2$ potential, \ie, $W=\lambda/x$. In that
particular
case, we have also a dynamical conformal symmetry, that of dilation, whose
generator is
$$D=-\textstyle{1\over2}xp+{i\over2}P_{(F)}-{i\over F}B_{(F)}, \e(24)$$
with $P_{(F)}$ given in $(15)$ and $B_{(F)}$ in $(6)$. $D$
satisfies the following commutation relations:
$$[D,Q]=-{i\over F}Q, \qq [D,H]=-iH. \e(25)$$
For $F=2$, we have $B_{(2)}=-{1\over2}+P_{(2)}$ and we recover the usual
dilation operator
$D=-{1\over2}xp+{i\over4}$.
In ordinary SUSY \qm, there are also other dynamical symmetries, namely, the
special conformal
($K$) and superconformal ($S$) ones. For $F>2$, this is not the
case\footnote{$\da$}
{Actually, $D'=-{1\over2}xp+{i\over4}$ and $K'={1\over2}x^2$ form an $SO(2,1)$
algebra with $H$ for any order $F$, but they have the wrong commutation
relations with $Q$ (for $F>2$) and therefore do not generate proper subalgebras
of the fractional Super-Virasoro algebra.}
since their
existence would imply an infinite conformal algebra.
To see this, remember first that a closed subalgebra of the (Neveu-Schwarz
sector of the) Super-Virasoro algebra is the $OSp(2,1)$ superalgebra
generated by the bosonic generators $H,D,K$ (\ie, $L_1,L_0,L_{-1}$ in the
usual notation of the Virasoro generators) and the fermionic ones $Q,S$ (\ie,
$G_{1/2},G_{-1/2}$).
[The bosonic operators alone form an $SO(2,1)$ algebra.]
In opposition,
the maximal finite closed subalgebra of the {\it fractional}\/
super-Virasoro algebra$^{[5]}$ (for $F>2$) is generated by $H,D$ and $Q$
(\ie, $L_1,L_0$ and $G_{1/F}$) only. As soon as  one adjoins $L_{-1}$ or
$G_{-1+1/F}$,
the entire infinite algebra follows.
For more details, see Refs.~[5].

\vskip 0.5cm
\noindent {\bf 4. Fractional supersymmetry transformations}
\vskip 0.2cm

In this section, we describe a way of constructing FSUSY
transformations within a pseudo-classical Lagrangian context, and we provide a
FSUSY invariant action for the free particle. The interacting case will be
discussed elsewhere.

Let us first recall the ordinary SUSY case. We work in one dimension. The
position of the particle is described by $x(t)$,
whereas its internal space is described
by a {\it real}\/ fermionic variable
$\psi(t)$ which satisfies $\psi^2=0$. The SUSY transformations are given by
$$\eqalignno{\delta x&=i\epsilon\,\psi &(26a)\cr
             \delta\psi&=\epsilon\,\dot x &(26b)\cr}$$
where $\epsilon$ is a fermionic infinitesimal parameter. Since
$\epsilon\psi=-\psi\epsilon$, the r.h.s of Eqs.~(26) are real. Note that
$\delta^2 x=i\epsilon_1\epsilon_2\,\dot x$ and $\delta^2
\psi=i\epsilon_1\epsilon_2\,\dot\psi$, so SUSY transformations are indeed the
square roots of time translations. An action invariant under (26) is
$$S=\int dt\;{1\over2}(\dot x^2+i\dot\psi\psi). \e(27)$$
More precisely, the Lagrangian within (27) varies by the total time derivative
$d[{i\over2}\dot x\psi]/dt$. Note that since $\psi\dot\psi=-\dot\psi\psi$, this
action is real.
The variables $x$ and $\psi$ can be viewed as the components of a superspace
coordinate $Z(t,\t)=x+\psi\,\t$
where $\t$ is a real grassmann variable, \ie, satisfying $\t^2=0$. We say that
$x$ and $\psi$ respectively belong to the sector-0 and sector-1 of the theory
($\t$ is also in sector-1 and sectors are defined modulo 2).

We now turn to the next order, i.e., {\it cube} roots of time translations.
We need a theory with three sectors. The bosonic variable $x(t)$ remains a
sector-0 quantity, but now we must have
{\it two}\/ types of {\it real}\/ internal-space variables, $\phi(t)$ and
$\psi(t)$, which respectively belong to sector-1 and sector-2. These three
fields can be viewed as the components of a
fractional superspace$^{[3]}$ coordinate of order $3$:
$\z(t,\t)=x+\psi\,\t+\phi\,\t^2$,
where $\t$ is a real \pg\ variable satisfying $\t^3=0$ (and belonging to the
sector-1). Sectors are defined modulo 3.
We take the following commutation relations between the new fields:
$$\psi\dot\phi=q\dot\phi\psi,
\qq \dot\psi\phi=q\phi\dot\psi                      \eqno(28)$$
where $q$ is a primitive cube root of unity, \ie, satisfies (3) with $F=3$.
The FSUSY transformations of order 3 are given by
$$\eqalignno{\delta x  &=i\epsilon\,\a(1-q)\,\psi       &(29a)\cr
             \delta\psi&=i\epsilon\,\a(1-q^2)\,\phi  &(29b)\cr
             \delta\phi&=\epsilon\,e\,\dot x  &(29c)\cr}$$
where $\a$ is a free (bosonic) constant and $e^{-1}=3\a^2$. Now, $\epsilon$ is
a sector-1 infinitesimal parameter which must satisfy
$$\epsilon x=x\epsilon, \qq
\epsilon\phi=q\phi\epsilon, \qq
\epsilon\psi=q^2\psi\epsilon. \e(30)$$
We easily see that $\delta^3=-\epsilon_1\epsilon_2\epsilon_3\,d/dt$ and so
Eqs.~(29) do represent cube roots of time translations.
The reality of the r.h.s of (29) follows from (30).
An action invariant under (29) is
$$S=\int dt\;{1\over2}\big[{\dot x}^2+i\beta(1-q)\dot\phi\psi
+i\beta(1-q^2)\dot\psi\phi\big] \eqno(31)$$
with $\beta=3\a^3$.
This Lagrangian is real, and varies by a total derivative under the
transformations (29) [see below].

Let us now turn to the general case, \ie, the $F^{\,\rm th}$ roots of time
translations with $F=1,2,...\,$. We need $F$ {\it real}\/ fields
$\psi_{(i)}(t)$ $[i=0,1,...,F-1]$ which belong to the sector-($F-i$), with
$\psi_{(0)}\equiv x(t)$. Sectors are defined modulo $F$. These fields can be
viewed as the components of a
fractional superspace$^{[3]}$ coordinate of order $F$:
$\z(t,\t)=\sum_{i=0}^{F-1}\psi_{(i)}\,\t^i=x(t)+
\sum_{i=1}^{F-1}\psi_{(i)}\,\t^i$,
where $\t$ is a real \pg\ variable satisfying $\t^F=0$ (and belonging to
sector-1).
We introduce the commutation relations,
$$\psi_{(i)}\psi_{(F-i)}=q^i\,\psi_{(F-i)}\psi_{(i)}, \e(32)$$
where $q$ is a primitive $F^{\,\rm th}$ root of unity, \ie, satisfies (3).
Taking the time-derivative of both sides of $(32)$ and using the symmetry under
$i\to F-i$, we get
$$\psi_{(i)}\dot\psi_{(F-i)}=q^i\,\dot\psi_{(F-i)}\psi_{(i)}. \e(33)$$
Although (32) might be consider more fundamental, only (33) is really needed
here (to ensure the reality of the action given below).
No other commutation relations, \ie\ between $\psi_{(i)}$ and $\psi_{(j)}$ with
$j\not=F-i$, are required.
For $F=2$ the relations $(32)$ and $(33)$ reduce [with the notation
$\psi\equiv\psi_{(1)}$] to the proper SUSY results $\psi^2=0$ and
$\psi\dot\psi=-\dot\psi\psi$. Note also that $\psi_{(0)}$ is indeed a commuting
variable.

The FSUSY transformations of order $F$ are given by $(i=1,2,...,F-1)$:
$$\eqalignno{&\delta\psi_{(i-1)}=i\epsilon\,\a(1-q^i)\,\psi_{(i)}&(34a)\cr
             &\delta\psi_{(F-1)}=\epsilon\,e\,\dot x  &(34b)\cr}$$
where $e^{-1}=F\a^{F-1}$ and where the infinitesimal parameter $\epsilon$
belongs as before to the sector-1. We now have $\delta^F
\psi_{(i)}=i^{F-1}\epsilon_1...\epsilon_F\,\dot\psi_{(i)}$, since
$\prod_{i=1}^{F-1}(1-q^i)=F$.
Our previous results (26) and (29) are recovered
for $F=2$ and $F=3$ (with the notation $\psi_{(1)}\equiv\psi$ and
$\psi_{(2)}\equiv\phi$). For $F=1$, we simply have $\delta x=\epsilon\,\dot x$.
To ensure that the r.h.s of the transformations $(34)$ are real (and that the
action given below is invariant), we must take the following commutation
relations between $\epsilon$ and $\psi_{(i)}$:
$$\epsilon\,\psi_{(i)}=q^{-i}\psi_{(i)}\epsilon.            \e(35)$$
An action invariant under (34) is:
$$S=\int dt\;{1\over2}\,\big[{\dot
x}^2+i\beta\sum_{i=0}^{F-1}(1-q^{-i})\dot\psi_{(i)}\psi_{(F-i)}\big] \e(36a)$$
with $\beta=F\a^F$. More precisely, the Lagrangian varies by the total
derivative
$$\delta L={d\over dt}\,{\textstyle{\epsilon\over2}}\!\!\left[i\a(1-q)\,\dot
x\,\psi_{(1)}+\dot x^2\,\delta_{F,1}\right]. \e(37)$$
Alternatively, using
$(33)$ and the symmetry of the action under the substitution $i\to F-i$, we may
rewrite $(36a)$ as
$$S=\int dt\;{1\over2}\,\big[{\dot
x}^2-i\beta\sum_{i=0}^{F-1}(1-q^{-i})\psi_{(i)}\dot\psi_{(F-i)}\big]. \e(36b)$$
Note that the action is real.
The first term of the sum in $(36)$ always vanishes, but is included because it
allows us to take the $F=1$ limit, which corresponds to the
spinless particle: $S=\int dt\,{1\over2}{\dot x}^2$.
The cases $F=2$ (with $\a=1/2$) and $F=3$ reduce to those given previously in
(27) and (31).

One may go further and construct the conserved N\"other charges associated with
these FSUSY transformations.$^{[3,7]}$ In Ref.~[3], we also give a
fractional-superspace formulation of the action (36), in order to make its
FSUSY invariance manifest.

\vskip 0.5cm
\noindent {\bf 5. Conclusion}
\vskip 0.2cm

We conclude with a general remark.
The SUSY algebra $Q^2=H$, the PSUSY one $Q^3=QH$ and the FSUSY one
$Q^3=H$ are concrete realizations of subalgebras of respectively the
superVirasoro, the
para-superVirasoro and fractional superVirasoro algebras$^{[4]}$.
Recently, these generalizations of the superVirasoro algebra have
been incorporated within a unified formalism, written in terms of fractional
superspace coordinates, under the name {\it generalized super-Virasoro
algebras}\/$^{[4]}$. This construction also contains new types of algebras
such as the following paragrassmann extention of the FSUSY algebra
of order 3: $Q^4=QH$ with $Q^3\not=0$. Quantum-mechanical realizations
of such algebras will be discussed elsewhere.


\vskip 0.5cm
\centerline{\bf Acknowledgments}
\vskip 0.2cm

I am pleased to thank K. Dienes and L. Vinet for
useful comments.
This work is supported in part by a fellowship from
the Natural Sciences and
Engineering Research Council (NSERC) of Canada.

\vfill\eject

\centerline{\bf References}
\vskip 0.2cm
\par
\frenchspacing

\item{[1]}
V.A. Rubakov and V.P. Spiridonov, {\it Mod. Phys. Lett.} {\bf A3}, 1337 (1988).

\item{[2]}
S. Durand, M. Mayrand, V.P. Spiridonov and L. Vinet, {\it Mod. Phys. Lett.}
{\bf A6}, 3163 (1991).

\item{[3]}
S. Durand, ``Fractional Superspace Formulation of Generalized Mechanics",
preprint McGill/93-06, May 1993 (hep-th/9305130),
to appear in {\it Mod. Phys. Lett.} {\bf A8}, No. 24 (1993).

\item{[4]}
S. Durand, ``Extended Fractional Supersymmetric Quantum Mechanics",
pre-print McGill/93-03, April 1993 (hep-th/9305129),
to appear in {\it Mod. Phys. Lett.} {\bf A8}, 1795 (1993).

\item{[5]}
S. Durand, {\it Mod. Phys. Lett.} {\bf A7}, 2905 (1992);
``Generalized Super-Virasoro Algebras and Fractional Superspace",
preprint McGill/92-43, to appear in the {\it Proceedings of the
XIX International Colloquium on Group Theoretical Methods in Physics},
Salamanca, Spain, July 1992 (CIEMAT publisher, 1993).

\item{[6]}
C. Ahn, D. Bernard and A. LeClair, {\it Nucl. Phys.} {\bf B346}, 409 (1990);
A.T. Filippov, A.P. Isaev and A.B. Kurdikov,
{\it Mod. Phys. Lett.} {\bf A7}, 2129 (1992);
``On Paragrassmann Differential Calculus", hep-th/9210075;
``Paragrassmann Extensions of the Virasoro Algebra", hep-th/9212157.

\item{[7]}
S. Durand, ``Fractional Supersymmetry",
preprint McGill/93-05 (March 19-93), to appear in the {\it Proceedings of the
VII J.A. Swieca Summer School on Particles and Fields},
S\~ao Paolo, Brazil, January 1993 (World Scientific, 1993).

\par
\vfill
\end